\documentstyle[12pt]{article}
\newcommand \dsty \displaystyle

\begin{document}\setcounter{page}{0}
\setlength{\baselineskip}{20pt}
\begin{center}
{\Large \bf University of Montpellier}
\hspace {1cm}
            {\bf PM/93-20}, June 93
\end{center}
\vspace{2cm}
%\vspace {1cm}\\
\begin{center}
%---------------------titre ---------------------------------------
{ \Large \sc
               Poking at the correspondence principle\\
              with derivative couplings\\
            }
%------------------------------------------------------------------
\vspace{1cm} {\large G. Moultaka}\footnote[0]
   {address after Oct. 93, SLAC, P.O.Box 4349, Stanford, CA. 94309.}
\vspace{.5cm}\\  Laboratoire de Physique Math\'{e}matique\\
Unit\'{e} Associ\'{e}e au CNRS n$^o$ 040768,\\
Universit\'{e} de Montpellier II Sciences et Techniques du Languedoc \\
Place E. Bataillon, Case 50 F-34095 Montpellier Cedex 5
\end{center}
\vspace {.5cm}
\begin{center}
{\bf Abstract} \\
\end{center}
It is shown that quantum mechanical systems obtained from a
heuristic path integral
quantization of lagrangians of the form
     $$ L= \sum_{n} \frac{1}{n!} f_n(q) \dot{q}^n $$
violate in general the correspondence principle, unless $L$ is not more
than quadratic in $\dot{q}$. The field theoretic counterpart of this
result and its consequence on models of interacting scalar and
vector fields and particularly the electroweak interactions,
are briefly discussed.
\newpage
\section{Introduction}
In the early days of the old quantum theory, the correspondence
principle played an essential role in finding the right theoretical
models that describe the atomic systems [1]. Later on, the final
formulation of quantum mechanics had to implement this
principle automatically in the quantization program.
This is realized through the
more than familiar substitution rules :
\begin{eqnarray}
q & \rightarrow &Q \nonumber \\
p & \rightarrow & \frac{\hbar}{i} P \nonumber \\
\label{e1} \end{eqnarray}
in the classical hamiltonian, with the precaution that one should
symmetrize properly in the operators $Q$ and $P$ in order to
construct a hermitian
hamiltonian. Equivalently one can use the path integral formulation
to get the transition amplitudes [2,3],
\begin{equation} <q; t \mid q'; t' > = {\cal N} \int\,\, [\,dp\,][\,dq\,]
\mbox{\Large e}^{\dsty \,\frac{i}{\hbar}\,\int_{t'}^{t}\,dt\,\,
(\,p \dot{q} - H[p,q] \, )} \label{e2} \end{equation}
with a choice of quantization of the form
\begin{equation}
<q'\mid H(P,Q) \mid q> = \int\, \frac{d\,p}{2 \pi \hbar}
\mbox{\Large e}^{i\,p\,(q-q')/\hbar}\,\, H(p, {\cal S}(q,q')\,)
\label{e3} \end{equation}
where ${\cal S}$ is any symmetric function verifying ${\cal S}(q,q)=q$.
The usual choice  ${\cal S}(q,q')= (q+q')/2$
is the so--called Wigner quantization.
Equation (\ref{e3}) guarantees the hermiticity
of the quantum hamiltonian as well as the correct continuum limit [3].
Here $q$ and $p$ denote respectively the quantum mechanical variables
and their conjugate momenta.\par \vskip .5cm
Relying on eq.(\ref{e2}) we will prove in the next section, that the
rule in eq.(\ref{e1}) together with eq.(\ref{e3}) can be insufficient
to ensure
a {\sl correspondence} between the classical and quantum systems
in the limit $\hbar \rightarrow 0$, and that this correspondence
should be viewed as intimately related to a specific class of
interactions\footnote{For definiteness, subsequent use of eq.[2] will
be made in the
euclidean metric ($i \rightarrow -1$).}.
The result generalizes easily to scalar field theory
(and with some more work, to vector fields), and thus selects a
particular class of universally allowed effective lagrangians.
Perhaps more importantly, it thus universally rejects a wide class of
interactions.
In section 3 we comment
briefly the result as well as its physical meaning and possible
implications, specifically in the case of electroweak
interactions.
\section{The proof}
In this section we give the main steps which constitute our proof of
the relevant constraints, while more technical details
will be displayed elsewhere.
We should however stress here that since the proof requires manipulation
of formal series, one should be very careful
that such a manipulation is not wildly biasing the result.\par
We start from a general classical lagrangian of the form
\begin{equation}
L= \sum_{n\geq 0} \frac{1}{n!} f_n(q) \dot{q}^n  \label{e4} \end{equation}
where the dots denote time derivatives and the sum can be infinite.
We will then have to determine the classical hamiltonian and
subsequently integrate over
the conjugate momenta p in eq.(\ref{e2}) to obtain the effective action
$S_{eff}$. It will turn out that $S_{eff}$ will
generally develop, in the limit $\hbar \rightarrow 0 $, an induced
classical term which was not present in the initial
classical lagrangian eq.(\ref{e4}). This inconsistency is avoided
only if $n \leq 2$. The latter condition will constitute the
selection criterion for the {\sl allowed} interactions.\\
The first difficulty one has to face in deriving the explicit form of the
 classical hamiltonian
\begin{equation}
H(p,q) = p \dot{q} - L(q,\dot{q}) \label{e5} \end{equation}
is obviously due to the relation between $\dot{q}$ and the conjugate
momentum $p$, namely
\begin{equation} p = \sum_{n \geq 0} \frac{1}{n !} f_{n+1} (q)
\dot{q}^n  \label{e6}
\end{equation}
Indeed one needs to invert eq.(\ref{e6}) to get $\dot{q} = \dot{q}(p)$,
which seems a priori a desperately ugly task!
Let us however write down a heuristic solution by inverting eq.(\ref{e6})
 through a formal iterative procedure. One gets
after a straightforward inspection
\begin{equation} \dot{q} = - \frac{1}{f_2}\,[ f_1 -p +
\circ_{\!\!{}_{\!\!\!\!{}_{{}_{i=0}}}}^{\!\!{}
^{\!\!\!\!{}^{{}_{\infty}}}}F
(\,\,\sum_{n \geq 2} (-1)^n \frac{1}{n !}\,[(f_1 - p)/f_2\,]^n\,f_{n+1}\,
\,) \,] \label{e7} \end{equation}
where the function $F$ is defined by
 $$ F(y)= \sum_{n \geq 2}\, (-1)^n \,\frac{f_{n+1}}{f_2^n} \sum_{p=0}^n
 \left ( \begin{array}{cc} n \\ p \end{array} \right)
 \,f_1^p \,y^{n-p} $$
with $f_n \equiv f_n (q)$
and $(\,\circ_{\!\!{}_{\!\!\!\!{}_{{}_{i=0}}}}^{\!\!{}^{\!\!\!\!{}^
{{}_{\infty}}}}F\,)$ stands for an iterative application of $F$,
{\sl i.e.} $F\circ F\circ F\cdots \circ F \cdots $ .\\
Upon use of eqs.(\ref{e5}--\ref{e7}) one then gets
\begin{equation} H[p,q] = -\frac{p}{f_2}{\cal H} - \sum_{n\geq 0}
\frac{1}{n !} (-1)^{n} \frac{f_n}{f_2^n} \,{\cal H}^n
\label{e8}. \end{equation}
where $${\cal H} =f_1 - p +
\circ_{\!\!{}_{\!\!\!\!{}_{{}_{i=0}}}}^{\!\!{}^
{\!\!\!\!{}^{{}_{\infty}}}}F(\,\,\sum_{m\geq 2} \,(p) \,\,)$$
and
$$\sum_{m\geq 2} \,(p) \equiv \sum_{m \geq 2} (-1)^m \frac{1}{m !}
\,[(f_1 - p)/f_2\,]^m\,f_{m+1}$$
$H(p,q)$ is thus obtained as a formal series in powers of $p$.
Furthermore,
injecting eq.(\ref{e8}) in eq.(\ref{e2}) and integrating out the
conjugate momentum $p$, one obtains the complete effective lagrangian
$ L^{eff}$ of the system under considertion. It is then possible to check
 whether the {\sl correspondence principle} is satisfied, by comparing
$L$ in eq.(\ref{e4}) to the behaviour of $L_{eff}$ in the limit $\hbar
\rightarrow o$. Now to integrate out $p$ explicitly is obvioulsy the
next hurdle to face. For this we make use of the familiar
Wick expansions, however in our case one
will be able to exponentiate
essential parts of the expansions, which will be sufficient to draw
definite conclusions. Actually these parts have the same structure
as those which are usually absent for instance in the Green's functions,
as they vanish when the external source coupled linearly
to the system is turned off. In the present case they are due to the
linear term in $p$ in the exponential, eq.(\ref{e2}),
and are evidently non--vanishing. To illustrate what is going on in a
rather simple way let us first retain only up to cubic terms
and write down, to fix the notation,
\begin{equation}
p \dot{q} - H(p,q) = a_0 + b\,p - \frac{1}{2}a \,\,p^2 + \frac{1}{3!}
c\,p^3 + \cdots \label{e9} \end{equation}

One then expands the exponential $\mbox{\Large e}^{\frac{1}{3!}{\dsty
c\,p^3}}$ in eq.(\ref{e2}) and integrates over $p$ for each term of
the expansion using the usual trick of successive differentiation with
respect $b/\hbar$. It is interesting to note that the
leading terms in powers of $b$ which are leading in the $\hbar$ expansion
 as well, have no symmetry factors and can thus be
completely re--exponentiated. Including all other terms one is lead to
the following result
\begin{equation}
\int  d\,p \,\mbox{\Large e}^{\frac{\dsty -1}{\dsty \hbar}{\dsty
( a_0 + b\,p - \frac{1}{2}a \,\,p^2 + \frac{1}{3!} c\,p^3)}}
= \frac{{\cal N}}{\sqrt{Det[a]}}
\mbox{\Large e}^{\frac{\dsty -1}{\dsty \hbar}
{\dsty (a_0 + \frac{\dsty 1}{\dsty 2} b\, a^{-1} \,b  +
 \frac{\dsty 1}{\dsty 3!} c\, (a^{-1}\,b)^3 +\hbar\,
 \Delta^{(3)}(\hbar) )}}
\label{e10}
\end{equation}
where
\begin{equation}
\Delta^{(3)}(\hbar) = -ln[1+\mbox{\Large e}^{\frac{\dsty 1}{\dsty
\hbar}\,{\dsty \frac{\dsty 1}{\dsty 3!} c\, (a^{-1}\,b)^3}}\,
\sum_{m \geq 1}
\sum_{p \geq 1}^{[3m/2]} \,\,\frac{(-1)^{p-m}}{m!}(\frac{\dsty 1}
{\dsty 3!} c)^m\, \hbar^{p-m}\, (a^{-1})^p (a^{-1} b)^{3m-2p}\, s_m^p]
\label{e11}
\end{equation}
and $ s_m^p = \frac{(3m)!}{2^p\,p! \, (3m-2p)!}$ is a symmetry factor
\footnote{Although not clearly explicited throughout, one should keep
in mind that $f_n(q)$ and $q$ can in general be respectively a tensor
of rank $n$ and a vector, in a given space of degrees of freedom.
Thus from
eq.(\ref{e12}), $a_0$ is a pure number, $b$ a vector,
$a= [f_2(q)]^{-1}$ a matrix and $c= f_3^{lmn} a_{li} a_{mj} a_{nk}$
a tensor of rank
three. Also all the products in the exponential in eq.(10) should be
understood in the functional sense with the integration
$\int dt$.}.
 Comparison of eqs.(\ref{e8}) and(\ref{e9}) gives
after the change of variable $p \rightarrow p- f_1$,
\begin{eqnarray}
a_0 & = & f_0(q) + f_1(q)\,\dot{q} \nonumber \\
b & = & \dot{q} \nonumber \\
a &= & \frac{1}{f_2(q)} \nonumber\\
c & = &  \frac{f_3(q)}{f_2(q)^3} \nonumber \\
\label{e12}
\end{eqnarray}
Inserting these values in eq.(\ref{e10}) yields finally the effective
lagrangian up to third order
\begin{equation}
L_{eff}^{(3)} = f_0(q) + f_1(q) \dot{q} + \frac{1}{2} f_2(q)
\dot{q}^2 + \frac{1}{3!} f_3(q) \dot{q}^3 + \hbar \Delta^{(3)}(\hbar)
\label{e13}
\end{equation}
One can actually obtain the full effective lagrangian iteratively along
the same lines, using as expansion parameter
$\epsilon = 1/ f_2(q)$, where it is assumed that $f_n(q)$ is of order
one, if $n\geq 3 $. This expansion is of course nothing but
a perturbation around the gaussian form. The general form of $L_{eff}$
to the $N^{th}$ order is found to be
\begin{equation}
L_{eff}^{(N)} = \sum_{n=0}^N \frac{1}{n!} f_n(q) \dot{q}^n + \hbar
\Delta(\hbar)
\label{e14}
\end{equation}
where $\Delta(\hbar)$ is now a much more complicated expression than in
eq.(\ref{e11}) and will not be displayed here.
Again eq.(\ref{e14}) can be obtainded by resumming the leading powers
in $b$ similarly to eq.(\ref{e10}), and shows clearly
that one does recover the classical lagrangian $L$ as part of the
effective largrangian (as it should). However, the question
we want to address now is whether $L$ is the full classical part of the
effective lagrangian, that is whether $\hbar \Delta(\hbar)$
vanishes in the limit $\hbar \rightarrow 0$. The answer to this question
requires technical but rather straightforward manipulations.
For the sake of simplicity we present this here only in the case where
$\Delta(\hbar)$ is given by eq.(\ref{e11}). Note that in
eq.(\ref{e11})
one cannot obtain directly the behaviour of $\Delta(\hbar)$ in the
classical limit, since in the argument of the log
the exponential increase
can be possibly compensated by the {\sl infinite} alternating power
series in $1/\hbar$.
We will actually prove that, as far as $c \neq 0$ the argument of the
log behaves
like $\mbox{\Large e}^{\frac{\dsty -1}{\dsty \hbar}}$ in the
quasistatic limit ($\dot{q}$ very small)
that is \\
{\sl $\hbar\Delta(\hbar)$ induces extra classical effects which are not
present in the initial classical lagrangian, unless $c = 0$}.
\\ As the proof is mainly
technical we exhibit here only the important steps.
The following identities will prove useful :
\begin{equation}
\frac{(3m)!}{(3m-2p)!} = \sum_{q=0}^{2p -1}\,\,\frac{(-1)^q}{q!}\,
(3m)^{(2p -q)}\,\,\sum_{i_1 \neq i_2 \neq i_3 \cdots \neq i_q \geq 1}
^{2p -1} i_1 i_2 i_3 \cdots i_q \label{e15} \end{equation}
and for any $X$,
\begin{equation}
\sum_{p =0}^{\infty} \,\, \frac{p^{\dsty s}}{p!} \,X^{\dsty s} =
(\sum_{{\dsty n}=1}^{\dsty s}\,\,
 \mbox{\Large a}_{\dsty s,n}\,X^{\dsty n})\,\mbox{\Large e}^X
 \label{e16} \end{equation}
where
\begin{eqnarray}
\mbox{\Large a}_{\dsty s,s}&=&1 \nonumber \\
\mbox{\Large a}_{\dsty s,n}&=& \sum_{k_1 \geq k_2 \geq k_3 \cdots
k_{s-n}\geq 1}^{n} k_1 k_2 k_3 \cdots k_{s-n} \,\,\,\,\,\,\,\,\,\,\,\,
(\mbox{for} \,\,\,n <s)
\label{e17} \end{eqnarray}
Now we rewrite eq.(\ref{e11}) as
\begin{equation}
\Delta(\hbar) = ln[1 + \mbox{\large e}^{\frac{\dsty X}{\dsty \hbar}}
\bar{\delta}(X,Y)]
\label{e18}
\end{equation}
with
\begin{equation}
\bar{\delta}(X,Y) = \sum_{m = 1}^{\infty} \sum_{p = 1}^{m-1} \,
(-\hbar)^{p-m} \frac{(3m!)}{m!\,(3m-2p)!\,2^p\,p!}\,X^m\,Y^p
\label{e19}
\end{equation}
and \begin{equation} X \equiv \frac{c}{3!} (a^{-1} b)^3\,\, , Y
\equiv a^{-1} (a^{-1}b)^{-2} \label{e19bis} \end{equation}
In eq.(\ref{e19}) we retained only terms which do not vanish in the
limit $\hbar \rightarrow 0$.
Using eq.(\ref{e15}) in eq.(\ref{e19}) one finds readily
\begin{equation} \sum_{p=1}^{\infty} \sum_{q=0}^{2p -1}
\sum_{m=p+1}^{\infty}\, (-\hbar)^{p-m}\, \frac{(-1)^q \, 3^{2p-q}}{
2^p\,p!\,q!}\,\frac{m^{2p-q}}{m!}\,X^m\,Y^p \,
\sum_{ i_1 \neq i_2 \neq i_3 \cdots \neq i_q \geq 1}^{2p -1}
 i_1 i_2 i_3 \cdots i_q  \label{e20} \end{equation}
Now we make use of eq.(\ref{e16}) in the form
$$\sum_{m =p+1}^{\infty} \,\, \frac{m^{2p-q}}{m!} \,X^{m}
= (\sum_{{\dsty n}=1}^{\dsty 2p -q}\,\,
 \mbox{\Large a}_{2p-q,n}\,X^{\dsty n})\,
 \mbox{\Large e}^X - \sum_{m =1}^{p} \,\, \frac{m^{2p-q}}{m!} \,X^{m} $$
to write
\begin{equation}
\bar{\delta}(X,Y) = \mbox{\large e}^{\frac{\dsty -X}{\dsty \hbar}}\,
\sum_{p=1}^{\infty} \sum_{q=0}^{p-1} \sum_{n=p+1}^{2p-q}
\,(-\hbar)^{p-n}\,\frac{(-1)^q 3^{2p-q}}{2^p\,p!\,q!}\,
X^n\,Y^p\,\mbox{\Large a}_{2p-q,n}\,
\!\!\!\sum_{i_1 \neq i_2 \cdots \neq i_q \geq 1}^{2p -1}\!\!\! i_1 i_2
\cdots i_q  \nonumber\\ \label{e21} \end{equation}
In the above equation we again dropped out terms which are irrelevant
in the limit $\hbar \rightarrow 0$ and thus made
consistently the replacement
$$  \sum_{p=1}^{\infty} \sum_{q=0}^{2p-1} \sum_{n=1}^{2p-q} \rightarrow
\sum_{p=1}^{\infty} \sum_{q=0}^{p-1} \sum_{n=p+1}^{2p-q}. $$
We are thus lead to the crucial fact that
the exponential suppression in the argument of the log in eq.(\ref{e11}),
 will be fully compensated by the exponential factor
spelled out in  eq.(\ref{e21}). One still has to study the behavior of
the remaining power series in $\hbar$. After a trivial change of variable
in the summation indices one gets for $\Delta(\hbar)$
\begin{equation}
\Delta(\hbar) = ln[ 1 + \sum_{l=1}^{\infty} \sum_{r=0}^{\infty}
\,(-1)^l \mbox{\Large (}\frac{9 X^2 Y}{2 \hbar}\mbox{\Large )}^l\,
\mbox{\Large (}\frac{9 X Y}{2}\mbox{\Large )}^r\,\,
\mbox{\large b}_{l,r} ]
\label{e22} \end{equation}
where $\mbox{\large b}_{l,r}$ denotes the following complicated but
well defined expression
\begin{equation} \frac{1}{(l+r)!} \sum_{q=0}^r \frac{(-1)^q}{3^q}\,
(\, \sum_{i_1 > i_2 >i_3> \cdots i_q =1}^{2l + 2r -1}i_1 i_2 i_3
\cdots i_q \,)\,
(\, \sum_{k_1 \geq k_2 \geq k_3 \cdots k_{r-q}
\geq 1}^{2l +r} k_1 k_2 k_3 \cdots k_{r-q}\,)
\label{e22bis} \end{equation}
Fortunately all what we need to remember from the above intricacy
is that the dependence in $\hbar$ comes exclusively
in the summation over $l$, {\sl i.e.} no $\hbar$ to the power $r$
and no $\hbar$ dependence in $\mbox{\large b}_{l,r}$.
It then follows that the argument of the log in eq.(\ref{e22})
behaves like $\mbox{\Large e}^{\dsty -9 X^2 Y/\dsty 2 \hbar}$
in the quasistatic limit, i.e. $\dot{q}$ sufficiently small.
To see this we note that
\begin{eqnarray}
1+\sum_{l=1}^{\infty} \sum_{r=0}^{\infty} \,\mbox{\Large (}
\frac{-9 X^2 Y}{2 \hbar}\mbox{\Large )}^l\,
\mbox{\Large (}\frac{9 X Y}{2}\mbox{\Large )}^r\,\,
\mbox{\large b}_{l,r} &=& \mbox{\Large e}^{\dsty -9 X^2 Y/2 \hbar} +
\sum_{l=1}^{\infty} \sum_{r=1}^{\infty} \,\mbox{\Large (}\frac{-9 X^2 Y}
{2 \hbar}\mbox{\Large )}^l\,
\mbox{\Large (}\frac{9 X Y}{2}\mbox{\Large )}^r\,\,
\mbox{\large b}_{l,r} \nonumber \\
&&\label{e23}
\end{eqnarray}
Now it is clear from eq.(20) that $XY$ depends linearly on $\dot{q}$
and thus
$\mbox{\Large (}\frac{9 X Y}{2}\mbox{\Large )}^r\,\,
\mbox{\large b}_{l,r}$ can be made arbitrarily smaller than $1$ for
sufficiently small $\dot{q}$.
In this limit the leading term in eq.(25) is the exponential and one gets
\begin{equation}
\hbar \Delta(\hbar)\,\,\,  \sim \,\,\, \frac{9}{2} X^2 Y \label{e266}
\end{equation}
which means that the effective lagrangian in eq.(\ref{e13}) is indeed
plagued with an induced ($\hbar$ independent )
classical contribution
which was not present in the original classical lagrangian eq.(\ref{e4})
(in the case $n\leq 3$). It is now clear from
eqs.(\ref{e12}, \ref{e19bis}, \ref{e266}) that
{\it the full correspondence between the classical and quantum system
will be restored only if $f_3(q) = 0$ for all $q$.} \par
This ends our proof. The generalization to the case $n \geq 3$ goes
along the same lines, albeit further technical intricacies,
and so will not be persued here\footnote{details will be given
in the appendix of ref.[14]}.
The general constraint is thus
\begin{equation} f_n(q) = 0 \,\,\,\,(n\geq 3 ) \label{e27}. \end{equation}
In the next section we discuss some of the
possible implications of the above result on issues relating to
low energy effective lagrangians in particle physics, and more
specifically to electroweak interactions.\\
\section{Comments and physical implications}
First, it is useful to note that the constraint eq.(\ref{e27}),
can be avoided (at least as far as the correspondence
principle is concerned) if the $f_n(q), \,\,(n\geq3)$ are themselves
quantum effects, that is if they depend on $\hbar$ and vanish
in the classical limit. This is typically what happens when the higher
derivative interactions are induced by
perturbative loop corrections.
Thus our result can be rephrased as follows :
{\sl if one insists on having higher derivative interactions,
then these are bound to be quantum corrections}.\par \vskip .5cm
All the above results can be carried over to relativistic field theory
almost straighforwardly, at least in the case of scalar fields to
start with\footnote{where now the dot in eq.(\ref{e4}) refers only to the
temporal derivative.}.
Supplemented by the requirement of Lorentz invariance one has from
eq.(\ref{e27}) (in the case of neutral scalar fields),
that all operators of the form
$$ \partial^{\mu_1} \phi\, \partial_{\mu_1} \phi \,\,
\partial^{\mu_2} \phi\, \partial_{\mu_2} \phi \cdots
\partial^{\mu_n} \phi\, \partial_{\mu_n} \phi \,\,\,\,
\,\,\,\,\,\,\,\, \mbox{with} \,\,\,\,\, n\geq 2$$
should be either absent or induced by radiative corrections. In the
issue this seems to tie up nicely with familiar
considerations related to renormalizability and the classification of
operators into relevant, marginal and irrelevant
in the infrared regime [4].\par \vskip .5cm
Yet one should stress two main differences. In deriving the usual
operator flow equations [4], the crucial
assumption is that the new physics is characterized by a sufficiently
high energy scale so that the infrared observables become
insensitive to almost any change in this scale. In the present approach
no specific reference whatsoever, to any underlying
physics is made, if not simply the general motivation for studying higher
derivative interactions.
The second point is that the flow equations analysis applies to any
type of operators (the main issue being the dimension of the
operators)[4], while in
the present case only those with temporal derivatives are
\mbox{concerned}.\par \vskip .5cm
Then it should be clear that the
resulting constraints will apply even if the low energy phenomena
were sensitive to a (near) new physics scale, and will thus lead to
complementary restrictions as regards the relevance of higher
dimensional operators. Specific examples will be given below. But before
doing so, a
further comment is perhaps useful at this stage, concerning the
case when the higher derivative operators are actually induced by quantum
effects. Even in this case one might still want the ensuing
interactions to be strictly vanishing when vector fields are
invovled, in order to preserve
the Lorentz invariance in the low energy regime. Indeed if eq.(\ref{e4})
denotes now a {\sl Lorentz invariant}
Lagrange density for an interacting covariant vector field
($ q \Rightarrow W_{\mu}$ and $\dot{q} \Rightarrow \partial_0 W_{\mu}$)
, then $\hbar \Delta(\hbar)$ is generally not a Lorentz invariant,
as can be inferred from the structure of eqs.(\ref{e266},
\ref{e19bis}, \ref{e12}) and should be formally required to vanish.
In fact this feature constitutes the full generalization of previous
investigations where only the
structure of the determinant in eq.(\ref{e10}) was considered [5,6].
So in some sense
                the analysis in this paper addresses the
question of consistent quantizability, from the point of view of the
correspondence between classical and quantum
systems or/and the preservation of space--time symmetries at the quantum
level. \par \vskip .5cm
We turn now to the subject of effective interactions in the context of
electroweak theory. This is of relevance to
various phenomenological studies carried out in the past few years [7],
to assess the ability of future colliders in testing
the gauge couplings among the electroweak vector bosons as predicted by
the standard model [8].
It also lead to some controversy [7, 9--11] related to whether
it is at all theoretically reasonable to expect "big'' deviations away
from the gauge couplings as a sign for "beyond'' the standard
model. Here we do not intend to discuss this very extensively.
We simply note that, stripped to its essence,
the answer to these questions
depends ultimately  on whether the supposed "new physics'' lies at
relatively low energy (comparable to the electroweak scale) or
at scales of the order of the Tev (or maybe $\sim 10^{15}$~Gev).
\par \vskip .5cm
Hereafter we want to show briefly how the analysis performed in this
paper can be used to give indirect hints about the possible physical
origin and expected orders of magnitude of the various effective
interactions studied in the literature. As an illustration we consider
the following higher dimensional operators taken from ref.[10].

\begin{eqnarray}
O_{B \Phi} & = & i B^{\mu\nu} (D_{\mu}\Phi)^{\dag} D_{\nu}\Phi \\
O_{W \Phi} & = & i(D_\mu\Phi)^{\dag} \stackrel{\rightarrow}{\tau}.
\stackrel{\rightarrow}{W}^{\mu\nu} D_\nu\Phi \\
O'_{W \Phi} & = & i( \Phi^{\dag}\stackrel{\rightarrow}{\tau}.
\stackrel{\rightarrow}{W}^{\mu\nu} \Phi)(D_{\mu}\Phi)^{\dag}
D_{\nu}\Phi \\
O_{\gamma\Phi}^{(1)} &=& \Box (\Phi^{\dag} \stackrel{\rightarrow}
{\tau}.\stackrel{\rightarrow}{W}^{\mu\nu} \Phi) D_\mu\Phi^{\dag}
D_\nu\Phi \\
O_{\gamma\Phi}^{(3)} &= &\partial_{\rho} (\Phi^{\dag}
\stackrel{\rightarrow}{\tau}.\stackrel{\rightarrow}{W}^{\mu\nu})
(D^{\mu} \Phi^{\dag} \stackrel{\rightarrow}{\tau}.
\stackrel{\rightarrow}{W}^{\mu\nu} \Phi
+ \Phi^{\dag} \stackrel{\rightarrow}{\tau}.
\stackrel{\rightarrow}{W}^{\mu\nu} D^{\mu} \Phi) \\
\nonumber
\end{eqnarray}
\begin{eqnarray}
\hat{\hat{O}}_Z &=& D_{\mu} ((D_{\nu} \Phi^{\dag}) \Phi - \Phi^{\dag}
D_{\nu} \Phi)\,[\, (D^{\mu} \Phi^{\dag}) D^{\nu}\Phi +
D^{\nu} \Phi^{\dag} D^{\mu}\Phi \nonumber\\
&& + \frac{1}{<\Phi>^2}((D^{\mu} \Phi^{\dag}) \Phi - \Phi^{\dag}
D^{\mu} \Phi)
 ((D^{\nu} \Phi^{\dag}) \Phi - \Phi^{\dag} D^{\nu} \Phi) \, ]\,  \\
\hat{\hat{O}}_{\gamma}^{(1)} &=& D_{\mu}D^{\rho}(\Phi^{\dag}
\stackrel{\rightarrow}{\tau}.\stackrel{\rightarrow}{W}^{\rho\nu} \Phi)
( D^{\mu} \Phi^{\dag} D^{\nu}\Phi +D^{\nu} \Phi^{\dag} D^{\mu}\Phi) \\
\hat{\hat{O}}_{\gamma}^{(2)} &=& D_{\mu}D^{\rho}(\Phi^{\dag}
\stackrel{\rightarrow}{\tau}.\stackrel{\rightarrow}{W}^{\rho\nu} \Phi)
((D^{\mu} \Phi^{\dag}) \Phi - \Phi^{\dag} D^{\mu} \Phi)((D^{\nu}
\Phi^{\dag}) \Phi - \Phi^{\dag} D^{\nu} \Phi)\nonumber\\
\end{eqnarray}
where $\Phi$ denotes the usual Higgs doublet and
$\stackrel{\rightarrow}{W}_{\mu}$ and $B_{\mu}$ are the $SU_L(2)$ and
$U_Y(1)$ gauge
bosons. The above operators have been studied in [10] as an
illustration of the possibility to generate anomalous W
 couplings from gauge invariant
higher dimensional interactions in the context of an effective
spontaneously broken model [12].
The question however, is to know whether these effects do not turn
out to be, after all, of the order of perturbatively
small radiative corrections, or even simply a reformulation of the
standard radiative corrections themselves, which
 generate small anomalous
couplings. To avoid this situation, one can try to interprete the
operators above as originating from quantum non--perturbative
effects. This could be achieved {\sl a priori} by assuming for
instance multi--fermion interactions at the underlying level, and making
use of general equivalence conditions {\sl \`{a} la}
Luri\'{e}--Macfarlane [13]. It is interesting to note that in
this case the induced
interactions become classical (at least in the leading $1/N$ expansion)
in the sense that $\hbar$ powers cancel out completely from
the vertices, once the poles in the $W$ and $\Phi$ propagators are
properly identified. In such a non--perturbative scenario the
magnitude of the induced
interactions might even be large.\par \vskip .5cm
Nevertheless the constraint eq.(29), can rule out this possiblity for
a certain type of operators. For instance $O_{\gamma\Phi}^{(1)}$
defined previously, has a contribution with $3$ time--derivatives of
$\Phi$ and should thus be suppressed according to eq.(29)
\footnote{ Note that
eq.(29) applies directly to this case even if $\Phi$ is complex, since
$\Phi$ will anyway induce\\ a neutral physical scalar.}.
An immediate consequence is that a large departure ($\delta_{\gamma}$)
from the $\gamma W^{+}W^{-}$ Yang--Mills coupling will be
difficult to induce from the gauge invariant combination
\begin{equation}
O_{\gamma\Phi} = \frac{i e}{M_W^2}\,( \frac{O_{\gamma\Phi}^{(1)}}
{<\Phi>^2} + \cdots)
\end{equation}
suggested in ref.[10], while small quantum effects are still allowed.
The same conclusion holds for $\hat{\hat{O}}_Z,
\hat{\hat{O}}_{\gamma}^{(1)}$ and $\hat{\hat{O}}_{\gamma}^{(2)}$.
Thus the CP--violating (C--violating, P--conserving) anomalous
$ZWW$ and $\gamma WW$ couplings, obtained respectively from
$\hat{\hat{O}}_Z$ and a special combination of
$\hat{\hat{O}}_{\gamma}^{(1)},
\hat{\hat{O}}_{\gamma}^{(2)}$ (and two other operators, see [10]
for details), should be vanishing or at most perturbatively small.
This is in accordance with what one would naively expect for
CP--violating effects in the bosonic sector.\par
\vskip .5cm
In contrast, the level to which we carried out the analysis in this
paper does not yet allow to draw definite conlusions concerning
other operators like $O_{B \Phi}, O_{W \Phi}, O'_{W \Phi}$,
$O_{\gamma \Phi}^{(3)}$ ... etc. As stated before, in this case one has
to take into account the quantization
of spin 1 (massive) fields, which generally brings up additional
constraints [6], related to the Loretnz invariance of the measure
in the path itegral. This is now under investigation.
\section{Conclusion}
To summarize, we have worked out in this paper the general path
integral quantization of higher derivative interactions in
the simplest quantum mechanical case. As far as we know this has never
been treated before beyond the quadratic case [3].
Stringent consistency requirements related to the correspondence
principle have thus been identified, at least in the quasistatic
limit and used to gain more
insight in the possible origin of higher dimensional effective
operators, involving scalar and vector fields in the context
of electroweak interactions.\\ The approach is however readily
generalizable to a variety of other physical situations,
and can be helpful in understanding the interplay between physics
at different energy scales.\par \vskip 0.5cm
{\sl Acknowledgements}\\
I thank J.L. Kneur for very many discussions of which Section 3
is but a very minor product and also for drawing my attention
to ref.[13].

\newpage
\begin{center} {\large\bf References} \end{center}
\begin{tabbing}
\=****\= \kill \\
\>[ 1] \> For a history see for instance ``Niels Bohr's Times''
by A. Pais,\\
\>     \> Clarendon Press, OXFORD 1991;\\
\>[ 2] \> E.S. Abers and B.W. Lee, Phys. Rep. 9C (1973) 1;\\
\>[ 3] \> See for instance chapters 2 and 3 of ``Quantum Field Theory
and Critical Phenomena'',\\
\>     \> by J.Zinn--Justin, Clarendon Press, OXFORD 1993;\\
\>[ 4] \> K.G. Wilson, Phys.Rev. B4 (1971) 3174, 3184;\\
\>     \> K.G.Wilson and J.G.Kogut, Phys. Rep. 12 (1974) 75;\\
\>     \> J. Polchinski, Nucl. Phys. B231 (1984) 269;\\
\>[ 5] \> T.D. Lee and C.N. Yang, Phys. Rev. 128, $n^o$2 (1962) 885;\\
\>     \> M. Nakamura, Prog. Theo. Phys, vol 33, $n^o$2 (1965) 279;\\
\>     \> K. H. Tzou, Nuovo Cim. 33 (1964) 286;\\
\>     \> S. Weinberg Phys. Rev. 138 $n^o$4 B (1965) 988;\\
\>     \> see also section 8 of chapter 10 of {\sl Gravitation \&
Cosmology} by S. Weinberg,\\
\>     \> Wiley (1972) and references therein;\\
\>     \> H. Aronson, Phys. Rev. 186 $n^o$5 (1969) 1434;\\
\>     \> (for a recent review see also, S. Peris, Ph.D. Thesis,
UAB--FT--213, May 1989);\\
\>[ 6] \> C. Latourre and G. Moultaka, Phys. Lett. B302 (1993) 245;\\
\>[ 7] \> K. Hagiwara, R.D. Peccei, D. Zeppenfeld and K. Hikasa, Nucl.
Phys. B282 (1987) 253;\\
\>     \> G.L. Kane, J. Vidal and C.--P. Yuan, Phys. Rev. D vol.39,
$n^o$9 (1989) 2617;\\
\>     \> F. Boudjema, K. Hagiwara, C. Hamzaoui and K. Numata,\\
\>     \> Phys. Rev. D 43 $n^o$7 (1991) 2223;\\
\>     \> A. de Rujula, M.B. Gavela, P. Hernandez and E. Masso,
Nucl.Phys. B384, (1992) 3;\\
\>     \> and corrected version June 1992;\\
\>     \> K.Hagiwara, S. Ishihara, R. Szalapski and D. Zeppenfeld,
Phys.Lett. B 283 (1992) 353;\\
\>     \> J. Layssac, G. Moultaka, F.M. Renard and G. Gounaris,
Int. J. Mod. Phys. A,\\
\>     \> Vol.8, No. 19, 1993;\\
\>     \> M. Bilenky, J.L. Kneur, F.M. Renard and D. Schildknecht,
BI-TP 92/44, PM 92-43,\\
\>     \> (to be published in Nulc. Phys.);\\
\>     \> G. Moultaka, talk at the $XIX^{th}$ International
Meeting on Fundamental Physics,\\
\>     \> ``Topics on Physics at High Energy Colliders'',
World Scientific, 1992,\\
\>     \> ed. E.Fernandez \& R.Pascual;\\
\>[ 8] \> S.L. Glashow, Nucl. Phys. 22 (1961) 579;\\
\>     \> S. Weinberg, Phys. Rev. Lett. 19 (1967) 1264;\\
\>     \> A. Salam, in : Proc. $8^{th}$ Nobel Symp., ed N. Svartholm\\
\>     \> (Almquist and Wiksell, Stockolm, 1968);\\
\>[ 9] \> C.P. Burgess and D. London, McGill-92/04 and 92/05;\\
\>[10] \> G. Gounaris and F.M. Renard, PM/92-31;\\
\>[11] \> M.B. Einhorn, UM-TH-93-12,Apr. 1993, to appear in the
proceedings of\\
\>     \> ``Unified Symmetries in the Small and in the Large'',
Coral Gables,FL, Jan. 93;\\
\>     \> M.J. Herrero, FTUM/92--06, Invited talk at the
International Workshop on \\
\>     \> ``Electroweak Symmetry Breaking'', Hiroshima,
Japan, Nov. 1991;\\
\>[12] \> W. Buchm\"uller and D. Wyler, Nucl. Phys. B268 (1986) 621;\\
\>[13] \> D. Lurie and A.J. Macfarlane, Phys. Rev. Vol.136, 3B, (1964),
816;\\
\>[14] \> G. Moultaka, PM-93-19 (in press);\\

\end{tabbing}

\end{document}